# Deep learning for channel estimation in FSO communication system

M. A. Amirabadi

Email: m_amirabadi@elec.iust.ac.ir

**Abstract-** Perfect channel estimation is very hard, time/ power consuming, and expensive; so it is not preferred (e.g. in mobile) communication systems. This paper seeks for new, cheap, low complexity, deep learning based solution. Several new combinations of deep learning and conventional structures (in different parts such as constellation shaper, channel estimator, and detector) are presented investigated, and compared over all atmospheric turbulence regimes from weak to strong (considering Gamma-Gamma atmospheric turbulence model). Results indicate that deep learning could provide close enough performance to the perfect channel estimation scheme, and it is immune to the atmospheric turbulence variation. The proposed deep learning based solutions are low cost, low complexity, with favorable performance. Accordingly, they are recommended for channel estimation in mobile communication systems. Because these system should deliver favorable, and cheap services to the costumers, which use a small mobile as transceiver that needs to be cheap, low complexity and low power consuming.

## I- Introduction

Free space optical (FSO) communication system, due to its advantages over conventional radio frequency systems, is one the promising technology for far future communication services [1]. FSO is very effective in outdoor communication links, because it uses lasers as the transmitter and accordingly could support higher range communications properly without the need for amplification or correction [2]. Despite the fact that in outdoor communication, eavesdropping is easier, FSO link is immunized to it [3]. However, aside many advantages, the atmospheric turbulence of the outdoor environment significantly degrades performance of FSO system, and limits its practical applications [4]. Accordingly, channel estimation could really help improving performance of FSO system, and making it reliable. However, for this purpose, it is required to transmit a pilot sequence and use a processing complexity.

Considering the complexity of the conventional communication systems, recently a new field of study emerged in optical communications, which by use of machine learning based algorithms tries to fulfil the cavity in complexity reduction of the existing techniques. Actually machine learning tries to remove the barriers by using the data itself. In the training phase, machine learns the structure of the data and finds the relation between input and output, then in the testing phase, the machine has the relation, so it could produce the desired output based on the input. The better training, the closer output to the desired. However, when the relation between input and output is very complex or when the input is not enough, it is required to learn the machine deeper and this is the idea of deep learning. Deep learning by consuming more complexity, could solve complex problems favorable.

Recently, many investigations considered DNN for fiber OC applications such combating the Fiber effects [5], Modulation Format Identification [6], Optical Performance Monitoring [7], Optical Amplifier Control [8], as well as some Optical Network applications [9]. However, there is no investigation over DNN for FSO communication. Considering machine learning applications in FSO, few basic investigations are developed in applications such as detection [10], distortion correction for sensor-less adaptive optics [11], and demodulator for a turbo-coded orbital angular momentum [12]. All of these works considered machine at the receiver side of a simple FSO-SISO system for a limited (specific) scenario; there is lack of comprehensive investigation in machine (deep) learning for FSO communication.

The purpose of this paper is to substitute deep learning based channel estimation by conventional high complexity and expensive scheme. To have a comprehensive view, this paper presents several new deep learning based FSO communication system. To the best of the knowledge, the novelties of this paper which are done for the first time in FSO system include

1) Presenting six novel combinations of deep learning and conventional FSO communication systems.

2) Use of deep learning in channel estimation, constellation shaping, and detection of FSO communication system.

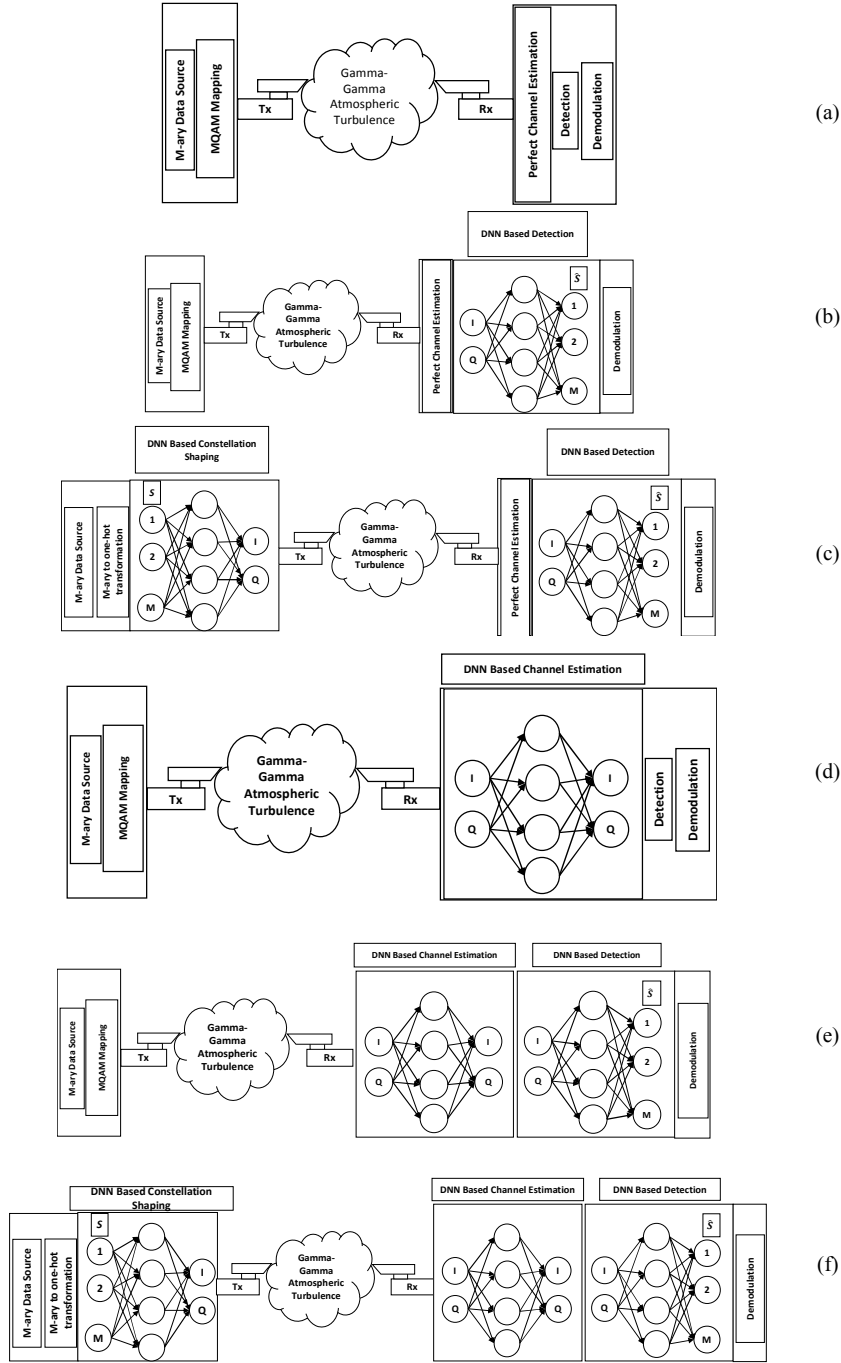

Fig.1. Three proposed DNN based FSO structures, a. QAM-perfect channel estimation-maximum likelihood, b. QAM-perfect channel estimation-DNN, c. DNN- perfect channel estimation-DNN, d. QAM-DNN-maximum likelihood, e. QAM-DNN-DNN, f. DNN-DNN-DNN

3) Comprehensively investigate deep learning effect in FSO communication system.

The rest of this paper is organized as follows; in section II system model is presented, section III is the results and discussions, section IV is conclusion of this work.

**II- System model**

As a core member in the machine learning community, deep learning (DL) showed significant performance in OC applications. The DL offers powerful statistical signal processing tools that could learn the received data impairments and generate an accurate probabilistic model for impairments. Among DL algorithms, DNN is the most widely used technique in OC, which has proved to be sufficient alternatives to conventional numerical, analytical, or empirical methods. DNN is simple and have low complexity, and can model complex multi-dimensional nonlinear relationships; it is generic and its response is fast. Due to these advantages, use of DNN in FSO for applications such as equalizer, constellation shaper, encoder, and detector, could significantly reduce complexity, cost, latency, and processing, while maintaining performance of the system. DNN first learns the model of the signal/system behaviour and can be used in high-level simulation and design, providing fast answers. DNN learns the input-output data relationship by using several hidden layers, which each consisted of multiple connected neurons by some weights, biases and activation functions that represent the importance of each connection. In this paper, several novel DNN based FSO communication channel estimators are proposed in which DNN is used. In the following, these structures are described in details.

### A. QAM-perfect channel estimation-maximum likelihood

As it is depicted in Fig. 1, a FSO system is considered where the information signal is transmitted by an optical transmitter and received by an optical detector. Coherent detection is assumed at the receiver side, therefore, the phase of the received signal can be detected. Considering $x$, as the transmitted symbol FSO optical transmitter, the received signal at the receiver aperture can be expressed as:

$$y = RIx + n, \qquad (1)$$

where $n$ is the receive aperture input additive white Gaussian noise (AWGN) with zero mean and variance $\sigma^2$; $I$ is the atmospheric turbulence intensity of the link between the FSO transmitter and receiver aperture, which is assumed to be Gamma-Gamma [13]; $R = \eta q/hf$ is photo detector responsibility [14, 15], $\eta$ is quantum efficiency of the photo detector, $q$ is the electron's charge, $h$ is Planck's constant, and $f$ is the optical frequency. The background noise limited receivers in which the shot noise created by background radiation is dominant compared to other noise components such as thermal noise, dark noise, and signal-dependent shot noise. Therefore, the noise term is modeled as signal-independent AWGN [16]. Assuming perfect channel estimation, the maximum likelihood receiver becomes as follows:

$$\hat{x}_u = \min_{\tilde{x}_u} |y - RI\tilde{x}_u|^2, \qquad (2)$$

where $\tilde{x}_u$ is a symbol of the transmitted constellation map.

### B. QAM-perfect channel estimation-DNN

In the second structure (Fig.1.b), QAM-perfect channel estimation-DNN, a DNN is used instead of the detector. The aim of this structure is to check what the benefit of using DNN as the detector is. Consider $x$ as the transmitted symbol, this symbol is first converted to a one-hot vector (because at the end, the output of the DNN would be a vector with size $M$, which is wanted to be the same as this one-hot vector), then mapped on an M-QAM constellation. Then the mapped symbol is transmitted from FSO transmit aperture. The transmitted signal is encountered by Gamma-Gamma atmospheric turbulence channel, and the receiver noise is added to the detected photocurrent of the photo detector. The received signal is entered a DNN with 2 input neurons (because the input is a complex number and the DNN can't afford complex numbers, so real and imaginary part would enter separately), $M$ output neurons, $N_{hid}$ hidden layers, $N_{neu}$ per layer neurons, $\alpha(.)$ activation function, $W$ weight, and $b$ bias. The purpose is to adjust DNN parameters (weight and bias) such that the receiver could better recover the original transmitted M-ary symbol.

In order to solve this problem efficiently, the DNN should be trained. The first step in training a DNN is selecting and tuning its hyperparameters [17], which include sample size to batch size ratio, layer type, number of layers, number of neurons, activation function, loss function, optimizer, learning rate, and number of iterations. Sample size to batch size ratio is important because entering the whole data at once into the DNN leads to underfitting while dividing it into several batches helps DNN to better understand the data structure. The number of layers, as well as neurons, should be adjusted by trial and test, and there is no specific rule for tuning them. For selecting the activation function, there is a long story, activation functions extended during the time and according to complexity, accuracy, and timing demands, there is a tradeoff between them, however some of them such as tanh, sigmoid, relu are shown to be proper for Machine Learning for OC applications, and frequently used in literatures.

After selecting and tuning the hyperparameters related to the DNN structure, it is turn to define hyperparameters directly related to train aspect. The inputs of each layer of DNN are multiplied by corresponding weights, added by biases, summed, and then passed the activation function. Outputs of each layer are the inputs of the next layer, and the procedure continues until reaching the last layer. Considering the original one-hot vector at the transmitter as $s$ and the output vector of the DNN as $\hat{s}$, the aim is to reduce the difference between $s$ and $\hat{s}$.

Therefore, a loss function should be defined and calculated for each individual transmitted symbol and expected over the whole batch size. The proposed loss function could be defined as [18]:

$$L(\boldsymbol{\theta}) = \frac{1}{K}\sum_{k=1}^{K}\left[l^{(k)}(\boldsymbol{s},\hat{\boldsymbol{s}})\right] \qquad (3)$$

where $\boldsymbol{\theta}$ is the DNN parameter vector (including weight and biase), $K$ is the batch size, $l(.,.)$ is loss function. The loss function we consider in this work is the cross-entropy, defined as [18]:

$$l(\boldsymbol{s},\hat{\boldsymbol{s}}) = -\sum_{i} s_i \log(\hat{s}_i). \qquad (4)$$

Several algorithms have been proposed to find good sets of parameters $\boldsymbol{\theta}$ which minimiozerd the proopsed loss function. One of the most popular algorithms is stichastic gradient desent (SGD) which obtain the parameters $\boldsymbol{\theta}$ iteratively as follows [18];

$$\boldsymbol{\theta}^{(m+1)} = \boldsymbol{\theta}^{(m)} - \eta \nabla_{\boldsymbol{\theta}} \tilde{L}(\boldsymbol{\theta}^{(m)}) \qquad (5)$$

where $\eta > 0$ is the learning rate, $m$ is the training step iteration, and $\nabla_{\boldsymbol{\theta}}\tilde{L}(.)$ is the estimate of the gradient. Actually, the error derivation ($\nabla_{\boldsymbol{\theta}}\tilde{L}(\boldsymbol{\theta}^{(j)})$) is fed back to the DNN as an updating guide; the positive step size is known as the learning rate ($\eta$). Optimization is a tricky subject, which depends on the input data quality and quantity, model size, and the contents of the weight matrices. Stochastic Gradient Descent methods could be used for determining update direction and solving (5) [18]. As the state of-the-art algorithm with enhanced convergence, the Adam algorithm is used for optimization during the training process in this work. All numerical results in the manuscript have been generated using the deep learning library TensorFlow [17].

### C. DNN- perfect channel estimation-DNN

The purpose of the third structure (Fig.1.c) is to investigate the effect of using DNN as joint detector and constellation shaper; this structure assumes that the perfect channel estimation is done. Consider $x$ as the generated M-ary symbol, it is first converted to a one-hot vector, then entered a DNN with $M$ input and 2 output neurons. For simplicity, and without loss of generality, other DNN structure is exactly the same as the DNN in Section II.A. Complex summation of the DNN output results in a complex number which stands for the location of the transmitted symbol in the signal constellation. Actually, this DNN is used for shaping the constellation to reduce the effect of atmospheric turbulence. Then the mapped symbol is transmitted, encountered by Gamma-Gamma atmospheric turbulence, and added by AWGN with zero mean and $\sigma^2$ variance. The received signal is entered a DNN exactly the same as the DNN of section II.B. The aim is to adjust the DNN parameters of the proposed structure simultaneously to reduce atmospheric turbulence effect, and recover signal better. The training procedure is exactly the same as descriptions of section II.B.

### D. QAM-DNN-maximum likelihood

The purpose of this structure (Fig.1.d) is to investigate the effect of using DNN as a channel estimator. Considering $x$, as the transmitted symbol FSO optical transmitter, the received signal at the receiver is entered a DNN with 2 input neurons, 2 output neurons, and $N_{hid}$ hidden layers, $N_{neu}$ per layer neurons, $\alpha(.)$ activation function, $W$ weight, and $b$ bias, actually the output of this DNN is the estimation of channel (which is done without any pilot symbols, and the channel is assumed to be un-correlated and stochastic). The received signal is entered a maximum likelihood detector, and by the use of the estimated channel, the transmitted signal is recovered. The aim is to adjust the DNN parameters of the proposed structure simultaneously to reduce atmospheric turbulence effect, and recover signal better. The training procedure is exactly the same as descriptions of section II.B.

### E. QAM-DNN-DNN

The purpose of this structure (Fig.1.e) is to investigate the effect of using DNN for joint channel estimation and detection. Consider $x$ as the transmitted symbol, this symbol is first converted to a one-hot vector (because at the end, the output of the DNN would be a vector with size $M$, which is wanted to be the same as this one-hot vector), then mapped on an M-QAM constellation. Then the mapped symbol is transmitted from FSO transmit aperture. The transmitted signal is encountered by Gamma-Gamma atmospheric turbulence channel, and the receiver noise is added to the detected photocurrent of the photo detector. The received signal is first entered a DNN with the same structure as section II.D, the output of this DNN is the channel estimation, then considering this channel estimation for removing the effect of channel, the signal (with removed channel effects) is entered a DNN with the same structure as section II.B. The aim is to adjust the DNN parameters of the proposed structure simultaneously to reduce atmospheric turbulence effect, and recover signal better. The training procedure is exactly the same as descriptions of section II.B.

Table.1. Tuned hyperparameters.

| Hyperparameter | Value |
|---|---|
| Modulation order | 16 |
| Number of layers | 4 |
| Number of hidden neurons | 40 |
| Batch size | 2^16 |
| Sample size/batch size | 4 |
| Number of Iterations | 1000 |
| Activation function | Relu |
| Loss | Softmax cross entropy |
| Optimizer | Adam |
| Learning rate | 0.005 |
| Gamma-Gamma Atmospheric turbulence Intensity | Strong ($\alpha = 4.2, \beta = 1.4$) <br> Moderate ($\alpha = 4, \beta = 1.9$) <br> Weak ($\alpha = 11.6, \beta = 10.1$) |
| Photo detector responsibility | R=1 |

### F.  DNN-DNN-DNN

The purpose of this structure (Fig.1.f) is to investigate the effect of using DNN for joint constellation shaping, channel estimation, and detection. Consider $x$ as the generated M-ary symbol, it is first converted to a one-hot vector, then entered a DNN with the same as the DNN in section II.C. Then the mapped symbol is transmitted, encountered by Gamma-Gamma atmospheric turbulence, and added by AWGN with zero mean and $\sigma^2$ variance. The received signal is first entered a DNN with the same structure as section II.D, the output of this DNN is the channel estimation, then considering this channel estimation for removing the effect of channel, the signal (with removed channel effects) is entered a DNN with the same structure as section II.B. The aim is to adjust the DNN parameters of the proposed structure simultaneously to reduce atmospheric turbulence effect, and recover signal better. The training procedure is exactly the same as descriptions of section II.B.

### III-   Results and Discussions

In this section, the simulation results of the proposed DNN based structures of part II are proposed and compared. Simulations for DNN based structures are done in Python/Tensorflow environment. The hyperparameters are tunned manually (Table.1.), and based on previous knowledge from literature. Considering FSO link in Gamma-Gamma atmospheric turbulence, strong ($\alpha = 4.2, \beta = 1.4$), moderate ($\alpha = 4, \beta = 1.9$), and weak ($\alpha = 11.6, \beta = 10.1$) regimes are considered in the simulations. The hyperparameter tuning here is done manually, but the proposed DNN based structures could achieve the performance of the state of the art conventional systems (in perfect channel estimation) or get close enough to the ideal results (in DNN channel estimator). Although hyperparameter tuning improves performance, the achieved improvements are not so much considerable that deserve adding complexity and processing to achieve them. In addition, it's complicated, and time and power consuming, so, the manual tuning here might not be a bad idea.

In Fig.2. symbol error rates (SER) of the proposed structures is plotted as a function of Es/N0 for a. weak, b. moderate, and c. strong atmospheric turbulence regime, when modulation order is $M = 16$. The aim of this paper is to investigate the effect of using DNN as a channel estimator at various system and channel models, and this aim is displayed in Fig.2. as can be seen, when perfect channel estimation is done, the effect of conventional and DNN based structures are the same, this is because when channel estimation is perfect, the problem that the DNN should solve is linear; actually, DNN outperformance over conventional systems is anywhere that model is not known or is non-linear. This shows that the proposed DNN based receiver does its work perfectly and efficiently in linear models. As can be seen, when channel estimation is not perfect, addition of each DNN system (detector, constellation shaper, and channel estimator), would improve performance of the system, because estimation of channel, when it is uncorrelated and stochastic is very hard without pilot symbol sequence, and this is exactly where DNN could be used. The performance difference between DNNs at each of these parts indicates that despite most of the applied investigations, which used DNN at the receiver side, DNN could be used as each parts of the communication system. Another thing that could be discussed is the difference between DNN based and conventional structures; as could be seen, this difference is almost the same for the all atmospheric turbulence regimes. So this is one of the advantages of the DNN based structures, immunity to the atmospheric turbulence variations makes this structure reliable. It is only tune and train it one time, it is expected to be robust at all atmospheric turbulence regimes. This reduces the cost and complexity required for running different systems for different atmospheric turbulence scenarios.

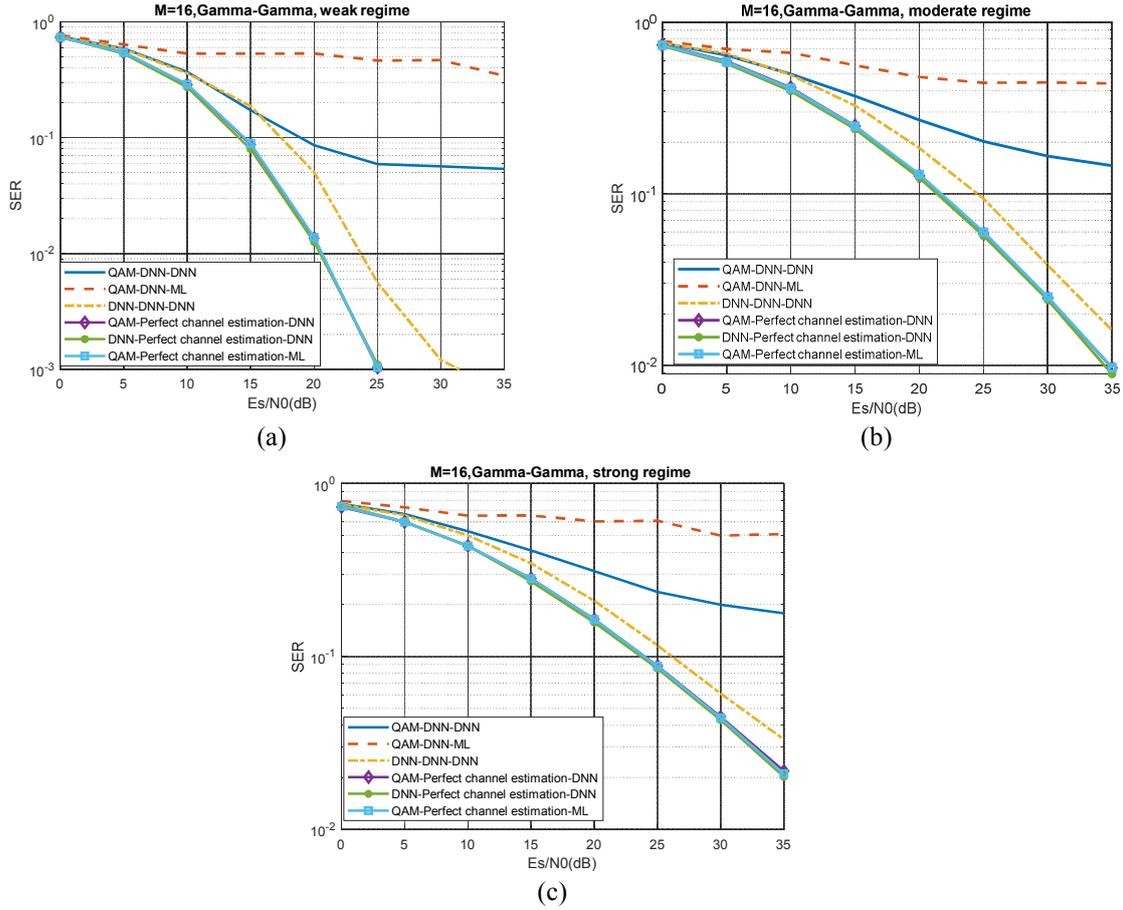

Fig.2. SER of the proposed structures as a function of Es/N0 for a. weak, b. moderate, and c. strong atmospheric turbulence regime, when modulation order is $M = 16$.

**IV- Conclusion**

For perfect free space optical (FSO) communication channel estimation, a pilot sequence should be transmitted and a complex processing should be applied; this technique is not user friendly (in mobile communications), because it is expensive and has lower data rate. The purpose of this paper is to seek for a low complexity, pilot independent solution (in deep learning). To have a comprehensive investigation, several combinations of deep learning and conventional structures are presented (as constellation shaper, channel estimator, and detector), and investigated at all atmospheric turbulence regimes from weak to strong. Results indicate that despite lower complexity deep learning based channel estimator, it could provide close enough performance to the perfect channel estimation scheme. In addition, it is shown that the deep learning based structure is immune to the atmospheric turbulence variation. According to the obtained results (e.g. low cost, low complexity, and favorable performance), deep learning is highly recommended for channel estimation in mobile communication systems. Because these system should deliver favorable, and cheap services to the costumers, in addition the costumer transceiver, which is a small mobile needs a low cost solution with lower complexity as well as power consumption.